# Force percolation of contractile active gels

José Alvarado[1,2,3], Michael Sheinman[4], Abhinav Sharma[4], Fred C. MacKintosh[4,5,6*], Gijsje H. Koenderink[1,*]

[1] Systems Biophysics Department, AMOLF, 1098 XG Amsterdam, The Netherlands

[2] Department of Mechanical Engineering, Massachusetts Institute of Technology, Cambridge MA 02139, USA

[3] Kavli Institute of Theoretical Physics, University of California, Santa Barbara CA 93106, USA

[4] Department of Physics and Astronomy, Vrije Universiteit, 1081 HV Amsterdam, The Netherlands

[5] Departments of Chemical & Biomolecular Engineering, Chemistry, and Physics & Astronomy, Rice University, Houston, TX 77005, USA

[6] Center for Theoretical Biological Physics, Rice University, Houston, TX 77030, USA

[*] Author to whom any correspondence should be addressed.

E-mail: g.koenderink@amolf.nl; fcmack@rice.edu

**Abstract**. Living systems provide a paradigmatic example of active soft matter. Cells and tissues comprise viscoelastic materials that exert forces and can actively change shape. This strikingly autonomous behavior is powered by the cytoskeleton, an active gel of semiflexible filaments, crosslinks, and molecular motors inside cells. Although individual motors are only a few nm in size and exert minute forces of a few pN, cells spatially integrate the activity of an ensemble of motors to produce larger contractile forces (~nN and greater) on cellular, tissue, and organismal length scales. Here we review experimental and theoretical studies on contractile active gels composed of actin filaments and myosin motors. Unlike other active soft matter systems, which tend to form ordered patterns, actin-myosin systems exhibit a generic tendency to contract. Experimental studies of reconstituted actin-myosin model systems have long suggested that a mechanical interplay between motor activity and the network's connectivity governs this contractile behavior. Recent theoretical models indicate that this interplay can be understood in terms of percolation models, extended to include effects of motor activity on the network connectivity. Based on concepts from percolation theory, we propose a state diagram that unites a large body of experimental observations. This framework provides valuable insights into the mechanisms that drive cellular shape changes and also provides design principles for synthetic active materials.

# 1 Introduction

Living cells constitute a highly unusual class of soft matter. Unlike most synthetic materials, cells are maintained in a state that is far from thermodynamic equilibrium by dissipative processes that convert chemical energy to mechanical work in the cytoplasm [1-3], plasma membrane [4], and nucleus [5,6]. The main origin of mechanical activity in the cytoplasm of plant and animal cells is the cytoskeleton, a space-spanning network of stiff protein filaments [7]. Two components of the cytoskeleton, filamentous actin (F-actin) and microtubules, are constantly remodeled by active (de-) polymerization. In addition, molecular motor proteins slide the filaments past one another or transport cargo across them. These processes turn the cytoskeleton into an active viscoelastic material. Strikingly, molecular motors are nanometer-sized and exert piconewton forces individually, yet ensembles of these motors can collectively drive large-scale events, allowing whole cells and tissues to move, change shape, and exert force. How can microscopic, molecular activity be coherently coordinated across longer length scales?

Similar questions have appeared also in the context of seemingly disparate systems such as cell colonies, suspensions of microscopic swimmers, flocks of birds, schools of fish, and animal herds [8]. Like cytoskeletal networks, these systems fall under a category of systems known as *active matter*. In active matter systems, individual units are driven by local energy sources. In the case of most flocking phenomena, individual units interact only locally with their neighbors, while long-range interactions are possible in cytoskeletal and extracellular networks, due to their filamentous constituents. When a large number of individuals simultaneously interact, correlations on length scales much longer than the size of an individual emerge, even when external fields or long-range interactions are absent [9,10]. *Active fluids* and *active gels* are subtypes of active matter distinct in their macroscopic viscoelastic properties. In both cases, the material is far from equilibrium due to internal driving by active microscopic agents. Fascinating patterns such as asters, spirals, vortices, or density waves have for instance been discovered in microtubule-kinesin suspensions [11,12], bacterial suspensions [13], and carpets of driven filaments [14,15]. Studying emergent phenomena in these active systems aids in understanding the mechanisms that drive complex biological processes such as mitotic spindle assembly [16] and bacterial colony formation [17]. At the same time, active systems provide inspiration to chemists for designing synthetic active materials [18,19].

Intriguingly, the actin-myosin cytoskeleton behaves rather differently from many active matter systems studied so far. Rather than exhibiting steady-state vortices or polarized asters, actin-myosin systems in cells contract as a result of stresses generated by the myosin motors inside mesh-like or bundled networks of actin filaments. In addition to steady contractile motion and cytoskeletal remodeling, such active stresses can also contribute to stochastic fluctuations, sometimes known as *active diffusion* [3,5]. Most animal cells possess a dense actin-myosin meshwork called the *cortex* that forms a thin (50–300 nm) layer anchored to the cell membrane [20,21]. The cortex mechanically protects the fragile cell membrane but also drives changes in cell shape [22]. Many large oocytes possess in addition a three-dimensional actin-myosin meshwork that pervades the cytoplasm and actively transports and organizes internal structures like chromosomes [23] and meiotic and mitotic spindles [24,25]. Cells that adhere to rigid extracellular environments have specialized contractile actomyosin structures referred to as *stress fibers*, which control cell stiffness and mechanosensitivity [26]. These contractile organelles exert nN-scale forces on their own [27], and when integrated across the whole cell give rise to µN-scale forces [28]. Cells themselves can use their actomyosin cytoskeleton to exert contractile forces on the surrounding extracellular matrix [29], thereby causing the entire tissue to contract [30-33].

It has been a long-standing question why actin-myosin networks are biased towards contraction. An important factor is the asymmetric force-extension response of actin filaments. Since actin filaments are semiflexible with a persistence length around 10 µm, they resist tension but readily buckle under compressive forces comparable to those generated by single molecular motors [34,35]. Experimental studies of reconstituted actin-myosin gels suggest that *network connectivity* is also a key parameter in biasing the gels towards contraction. Various studies showed that crosslinking of filaments allows myosin motors to propagate contractile stresses across system length scales [36-44].

Theoretical models have been developed on different scales to predict patterning and contractility in active systems. On the microscopic scale, early numerical simulations have predicted a variety of ordered steady states in microtubule-kinesin systems [11,45,46]. In addition, models that explicitly describe myosin-mediated sliding between actin filaments also predict contraction in filament bundles and networks [34,35,43,47-50]. On the continuum scale, models based on linear hydrodynamic equations describing liquid crystals supplemented with active driving have succeeded in predicting ordered aster and vortex patterns [51] and propagating waves [52,53]. Similar models also predict contraction (or *density instabilities*) in filament bundles [54,55] and isotropic crosslinked gels [53,56,57]. To account for the role of connectivity in biasing active networks towards contraction, a class of network models aimed at length scales between the microscopic and continuum have been developed. These have been inspired by much earlier work on marginal mechanical stability of networks [58] and concepts from percolation theory [42,59-63]. Different from passive networks, connectivity in active networks is not fixed but influenced by the internal activity. Stresses applied by motors affect the binding affinity of crosslinks [64,65] and thus connectivity. Experiments on reconstituted networks showed that motors can also reduce connectivity by severing [41,66] or depolymerizing actin filaments [67]. On the other hand, theoretical studies predict that motor activity mechanically stabilizes low-connectivity networks [61,62,68], consistent with experimental observations of cells showing that nonmuscle myosin-II contraction of cytoplasmic actin filaments is necessary to establish a stable cytoskeletal network [69]. Percolation models extended to include reciprocal feedback between connectivity and motor activity provide an interesting new approach to combine a continuum description of active networks with a microscopic description of the internal active driving.

Here we review studies on contractile active gels focused on reconstituted model systems based on actin, myosin, and crosslink proteins. Throughout this review, we will use the term *contractile active gel* to refer specifically to a mesh-like polymer network that contracts in response to molecular motor activity. This definition may apply both to molecular motors contracting intracellular polymers, as well as entire cells contracting extracellular polymers. We propose a state diagram based on previous experimental and theoretical studies of contractile active gels combined with percolation models of filamentous networks. The diagram identifies four main regimes of behavior in terms of two physical parameters: network connectivity and motor activity. We furthermore review experimental studies that show how these physical parameters can be tuned at the molecular level. The phase diagram we propose provides a broad framework that unites the seemingly disparate behaviors observed in experiments on different active systems: collective swarming, coarsening, cluster formation, tension generation, and contraction over varying length scales. It may help to classify and understand contractile properties of the actomyosin cortex in cells and developing embryos, and provide guidelines for designing synthetic active materials with desired macroscopic physical properties.

# 2 Experimental model systems for contractile active gels

How do cellular actin-myosin networks contract? To answer this question, researchers have established biomimetic model systems composed of purified actin filaments, myosin motors, and crosslink proteins *in-vitro* (Fig **1**). The advantage of this reductionist approach is that the biochemical composition of these simplified systems can be systematically controlled, allowing for direct quantitative comparison with physical models. Moreover, minimal model systems are useful for identifying the principles that are necessary as well as sufficient for networks to be contractile. In this section we briefly review the molecular components of contractile active gels.

## 2.1 Entangled actin networks

Actin filaments ("F-actin") are composed of two linear strands of globular actin subunits ("G-actin") that twist around each other to form helical filaments with a 37-nm pitch [70]. The G-actin monomers are comprised of two domains separated by a cleft that binds a divalent cation and either adenosine triphosphate (ATP) or adenosine diphosphate (ADP). Actin filaments are structurally polar because the monomers assemble head-to-tail with the ligand-binding clefts all directed towards one end (denoted the *minus end*, while the other end is called the *plus end*). Myosin II motors take advantage of the structural polarity to move in a directional manner toward the plus end. Moreover, hydrolysis of the ATP bound to G-actin monomers that add onto the plus end of a growing filament provides chemical energy that maintains different monomer on- and off-rates at the two filament ends. The plus end has a higher on-rate than the minus end, leading to a phenomenon called *treadmilling* that allows actin filaments to exert polymerization forces to drive cell migration. For purified actin solutions the treadmilling process is exceedingly slow, but *in vivo* the turnover of actin filaments is enhanced by several orders of magnitude by several proteins such as gelsolin and ADF/cofilin [71].

Due to its supramolecular architecture, F-actin is much stiffer than conventional synthetic polymers. Its thermal persistence length is close to 10 μm and thus comparable to its contour length, while being three orders of magnitude larger than its diameter of 7 nm [72]. As a consequence, F-actin filaments form space-filling networks already at volume fractions of less than 1% [73]. F-actin reconstituted from purified actin has an exponential length distribution with a typical filament length of 15 μm [74]. Actin filaments are generally thought to be much shorter in cells. Measurements on cortical actin in mammalian BSC-1 cells suggested lengths of ~3 μm [75,76]. The length of actin filaments in reconstituted gels can be controlled by adding physiologically relevant proteins that nucleate and/or sever filaments, such as formins, Arp2/3, gelsolin, and cofilin [77].

Above an actin concentration, $c$, of 0.1 mg/mL, actin filaments entangle and form semi-dilute mesh-like networks. At 1 mg/mL, the mesh has an average pore size of ~0.3 μm [78]. Filament entanglements govern the mechanical properties of semidilute actin solutions [73,79]. At intermediate timescales, entangled solutions behave like soft solids, with an elastic shear modulus $G'$ of less than 1 Pa (~100-fold softer than yogurt [80]), which exceeds the viscous shear modulus $G''$ by about 4-fold. On longer timescales, entangled solutions are fluid ($G''$ is larger than $G'$), because the entanglement constraints are eventually released by diffusion of the filaments along their contour ("reptation") [81]. Reconstituted F-actin solutions relax after ~10–100 min, depending on filament length [73]. This timescale is much longer than the stress relaxation time for actomyosin networks *in-vivo*, which is on the order of a few seconds primarily due to rapid F-actin turnover [82-85]. At short timescales, below 1 s, $G''$ is also larger than $G'$, and both moduli exhibit power-law dependencies on the deformation frequency. Theoretical models of wormlike chains predict that stress relaxation at short times is governed by transverse thermal bending fluctuations of the filaments, which lead to a $\omega^{3/4}$-dependence of the rheology [79,86], as observed in experiment [87-90]. For entangled solutions, an additional $\omega^{5/4}$-regime was predicted due to axial tension propagation [79,91,92], which was also validated experimentally [93].

## 2.2 Crosslinks

Cells can modulate the elastic properties and spatial organization of their actin cytoskeleton by cross-linking the filaments with dedicated crosslink proteins. Crosslink proteins usually have two actin-binding domains connected by a linker domain. The most common actin-binding domain is the calponin-homology domain, which is found across a broad class of crosslink proteins, including spectrin, filamin, fimbrin, and α-actinin [94]. These crosslinks are all homodimeric. Fascin proteins are unusual: they are monomers with two actin-binding domains, and bind actin filaments through β-trefoil domains [95]. Actin-binding crosslinks provide greater mechanical stability than the physical constraints from filament entanglements. As a result, the storage modulus $G'$ of crosslinked actin gels over a broad frequency range can be ~100x greater than entangled filament solutions. Crosslinked actin networks begin to stiffen with increasing molar ratio of crosslink to actin monomers, $R_X$, above a certain critical crosslink concentration. Furthermore, the critical concentration does not vary strongly with crosslink type (Table 1). These observations are consistent with percolation models of chemically crosslinked polymers [96] (see §3.2 below). However, the mechanical response of crosslinked actin networks exhibits several distinguishing characteristics.

First, the architecture and mechanical properties of actin networks are sensitive to the size and geometry of the crosslink. Fascin and fimbrin are compact, globular proteins that prefer to bind to tightly apposed filaments under a small angle; as a result, they usually generate tight, unipolar bundles [97,98]. In contrast, larger, rod-like crosslinks such as α-actinin form actin bundles of mixed polarity [98]. Large, fork-like linkers such as filamin can bind actin filaments over a wide range of angles, forming isotropic networks at low crosslink density and mixed network/bundle phases at high crosslink density [99]. However, the crosslink geometry is not always predictive of network architecture. The kinetics of actin polymerization and crosslink binding can sometimes dominate the final network structure [100-103]. Many crosslink proteins, whether rigid or flexible, tend to form actin bundle networks at sufficiently high crosslink concentrations. Bundling can stiffen actin networks compared to isotropically crosslinked networks. However, softening can also occur because bundled networks tend to deform in a more nonuniform (non-affine) manner [104], and also due to sliding [105] and clustering [106] of the bundles.

Second, the mechanical compliance of the crosslink proteins also strongly influences the mechanical response of crosslinked actin networks [107]. Rigid proteins such as scruin do not significantly deform when stressed. Thus, the entropic force-extension behavior of segments of actin filaments between crosslink points governs the elastic modulus of networks with rigid crosslinks [108,109]. Consequently, the elastic modulus is highly sensitive to crosslinking, with $G'$ varying over many orders of magnitude with changes in crosslink or actin concentration [110,111]. Moreover, due to the non-linearity in the entropic spring constant of actin filaments at high extensions, the networks stiffen at high stresses, a phenomenon known as stress stiffening [110,112]. In contrast, the elastic modulus of actin networks crosslinked with flexible proteins such as filamin is dominated by the compliance of the crosslinker. In this case, the elastic modulus is small when the network is subjected to small stress, but increases strongly once the crosslinks are fully stretched [113-115].

Third, crosslink proteins have a finite binding affinity for actin: they bind transiently with typical dissociation constants in the range of 0.1–3 μM [105,116-121]. This corresponds to binding free energies of 32–42 kJ mol[-1], or 13–17 times the thermal energy $k_B T$ at room temperature and typical crosslink unbinding times in the range of 1-10 s [98]. One notable exception is the acrosomal protein scruin [110], which has a much higher affinity. The molecular binding kinetics of actin crosslinks influences both the elastic and the viscous properties of actin networks. At timescales longer than the crosslink unbinding time, crosslink dissociation leads to stress relaxation, as seen for crosslinking by heavy meromyosin [122,123]. Theory predicts a non-single-exponential relaxation and corresponding viscoelastic response for times longer than the unbinding time, consistent with experiments with alpha-actinin crosslinks [124]. Recent simulations also predict that in case of bundled actin networks, transient crosslinker binding introduces various new rheological regimes at high, intermediate, and low frequencies [125]. Moreover, experiments revealed glass-like aging in actin networks bundled with fascin [126].

The binding kinetics of the crosslinks also influences the nonlinear response of actin networks to large stresses. Usually, tensile loads accelerate crosslink unbinding [64]. Such crosslinks are known as *slip bonds*. As a consequence, the network response becomes rate-dependent, with stiffening at high deformation rates and softening at small rates [127]. Typical rupture forces for actin crosslinks measured by single-molecule experiments with optical tweezers are in the range of 40-80 pN [128]. However, many biological adhesion molecules exhibit a surprising behavior known as *catch bond* behavior, whereby mechanical loads up to a certain force enhance the binding affinity for their ligand [65,129,130]. At even higher forces, a transition to slip bond behavior occurs. The crosslink α-actinin 4 is thought to exhibit such catch bond behavior. Structural analysis showed that this protein exhibits different stable conformations [131] and that mechanical forces can expose cryptic actin binding domains, thus enhancing the binding affinity for actin [132]. This catch bond behavior at the molecular scale translates into counterintuitive rheological properties on the network scale: actin networks crosslinked with α-actinin 4 shear-thicken and an applied shear stress extends the regime of solid-like behavior ($G' > G''$) down to lower frequencies [133]. Similar behavior was found for networks crosslinked by inactivated nonmuscle myosin IIB [134].

## 2.3 Myosin molecular motors

The myosin superfamily encompasses seventeen different classes that are each specialized for different cellular tasks (reviewed in [135]). The motors of the myosin II class (*conventional myosins*) are largely responsible for cell contractility. Although myosins within this class differ in their enzymatic and self-assembly properties [136,137], they share a common structural design consisting of two globular head domains joined by a long tail domain. The head domains bind to actin filaments and move towards the plus end using energy released from ATP hydrolysis, while the tail domains serve to assemble myosin molecules into bipolar filaments. Myosin filaments in muscle, or *thick filaments*, are longer than those in the cytoskeleton, which are called *minifilaments* (see §5.1 for a more detailed discussion). The bipolar structure of myosin filaments, with motor heads on the two ends and the tails packed in the center, allows myosin filaments to slide anti-parallel actin filaments in opposing directions. When embedded within a crosslinked actin meshwork, myosin bipolar filaments can thus be modeled as contractile force dipoles [138]. The forces of myosin motors in actin networks introduce nonequilibrium fluctuations that violate the fluctuation dissipation theorem [139], invalidating passive microrheology methods based on thermal fluctuations and requiring active microrheology to measure stiffness in cells [1,140,141]. Such nonequilibrium fluctuations can also introduce non-zero currents in the phase space of a system that are forbidden in thermal equilibrium [142,143].

Although individual bipolar filaments can exert both contractile and extensile forces, cellular and also reconstituted actomyosin networks tend to be contractile. In skeletal muscle, the origin of this asymmetry clearly lies in the arrangement of the actin and myosin filaments into a periodic and aligned array of so-called sarcomeres. Sarcomeres are repeating linear arrays of myosin thick filaments that are co-aligned with two antiparallel sets of actin filaments that have their minus ends in the center and their plus ends outwards and anchored at the Z-discs. The sliding motion generated by the myosin thick filaments thus leads to uniform contraction. A similar ordered arrangement but with varied polarity patterns is present in stress fibers in non-muscle cells [144].

In disordered networks such as the actomyosin cortex or the bulk actomyosin networks in oocytes, the origin of the asymmetry which favors net contraction is unclear since contractile arrangements are as likely as extensile ones. A range of mechanisms has been proposed. The mechanism that is best supported by experiments is one that attributes contractility to the nonlinear, asymmetric mechanics of actin filaments. Actin filaments readily buckle under compression [145,146] whereas they strongly resist stretching [147]. Experimentally, buckling of actin filaments was observed during contraction of quasi-2D actin-myosin networks and a correlation was observed between the macroscopic deformation and the amount of deformation of individual filaments [41,148]. Several theoretical studies argued the importance of actin-filament buckling in contraction [34,35,40,57,60,63,149-152]. However, other mechanisms have been shown to lead to contraction in the absence of buckling. One model proposed that contractile forces are generated by a "plucking" mechanism, where motors excite transverse fluctuations in an entangled meshwork [47]. It is not even strictly necessary to invoke filament deformation to explain contractility in filament-motor mixtures. If the motors transiently stall when they reach the plus end of actin filaments, contractility is favored in bundles as well as networks [54,55]. This mechanism appears to underlie recent reports of contractility in microtubule-motor systems [44,153]. When the motors themselves are modeled as finite-sized and deformable, contractility also naturally arises because the myosin minifilaments may move directionally along actin filaments, toward low-energy contractile configurations [47,59]. Steric repulsion between actively driven hard rods [154] or attraction due to entropic forces [155-157] can also result in contractile behavior.

# 3  Percolation models and marginal stability of passive systems

Both experimental [36-42,148,158] and theoretical studies [59-61] have demonstrated that connectivity provided by crosslinks is essential to contractility in disordered networks. To understand the role of connectivity in active systems, we first provide an introduction to marginal stability and percolation models of passive systems. These models describe how the microscopic connectivity of the network governs macroscopic material response properties and transitions between response regimes.

## 3.1  Conductivity percolation

Percolation models (reviewed in [159]) represent random networks by a lattice of points called *nodes*. Many kinds of lattices are possible, including square, triangular (shown in Fig. **2**), or honeycomb in 2D; as well as cubic, body-centered cubic, face-centered cubic, and diamond in 3D; or hypercubic at higher dimensionality. Models may consider *bonds*, or lines between nearest-neighbor nodes of a lattice. The *connection probability p*, which can range between 0 and 1, determines whether sites or bonds are occupied by a connection. These connections represent conduits in many kinds of transport problems: electrical (or thermal) current through random resistor networks [160,161], fluid flow through porous media [162], vehicle traffic flow [163], and forest-fire propagation [164].

How does $p$ determine a system's macroscopic conduction properties? As $p \to 0$, systems comprise disjointed *clusters*, or groups of adjacent connections (Fig. **2**a,b). Clusters have an exponential size distribution $P(s) \sim e^{-s/S}$ with $s$ cluster mass and $S$ typical cluster mass (both in units of number of sites or bonds). For the example of a linear 1D lattice, $S = \frac{1+p}{1-p}$. If a voltage is applied to two opposing ends of a random resistor network with $p \to 0$, current will not flow because the typical cluster diameter $l$ is much smaller than the system length scale $L$. The system thus behaves like an electrical insulator, with conductivity $\Sigma = 0$. As $p$ grows, clusters become larger (Fig. **2**c). In the limit $p \to 1$, the system comprises one globally connected, system-spanning cluster (Fig. **2**d) with diameter $l \sim L$. A network of resistors with $p \to 1$ responds as an electrical conductor, with conductivity $\Sigma$ proportional to the conductivity $\Sigma_0$ of one resistor.

Based on these two limits, one expects a transition from insulating to conducting states. Indeed, percolation models show that there exists a specific value $p_C$, called the *conductivity percolation threshold* (Table 2), where just enough connections form to allow one spanning cluster with $l \sim L$ (Fig. **2**c, red cluster). Clusters at $p_C$ have *fractal dimension* $d_f$ defined as $s \sim l^{d_f}$. This fractal shape implies that clusters have holes of various, scale-free sizes, and larger clusters surround smaller clusters called *enclaves* (Fig. **2**c, light green clusters). The size distribution of clusters $P(s) \sim s^{-\tau}$ exhibits a power law, with $\tau$ the *Fisher exponent* [165].  The Fisher exponent relates to the fractal dimension via the hyperscaling relation $\tau = \frac{d}{d_f} + 1$. For $p > p_C$, the spanning cluster becomes Euclidean ($d_f$ equal to the dimensionality of the system), as the added connections incorporate enclaves into the spanning cluster.

The spanning cluster's fractal shape implies that (i) the shortest path across the system (*conduction backbone*) meanders at $p_C$ and is much longer than $L$; (ii) the conductivity $\Sigma$ of a random resistor network just above $p_C$ is finite but small; and (iii) the characteristic timescales associated with charge transport diverge as $p \to p_C$ (*critical slowing down*). As $p$ increases away from $p_C$, the added bonds shorten the conduction backbone and $\Sigma$ increases toward $\Sigma_0$. One can show that $\Sigma \sim \Sigma_0 (p - p_C)^{f_C}$, with $f_C$ the *conductivity exponent*. This relation was experimentally verified in semiconductor sheets with punched holes [166].

The conductivity threshold represents a continuous second-order phase transition. The average distance $\xi$ of two sites belonging to the same cluster acts as a correlation length. One can show that $\xi$ diverges at $p_C$ with $\xi \sim |p - p_C|^{-\nu_C}$. The mass $S'$ of all clusters except for the largest acts as a susceptibility, and also diverges at $p_C$ with $S' \sim |p - p_C|^{-\gamma}$, indicating critical behavior similar to thermal transitions. Critical exponents like $\tau$, $f$, $d_f$, $\nu$, and $\gamma$ depend only on the dimensionality of the system (and not on the lattice geometry used, or whether sites or bonds are considered) and indicate the *universality class* of the transition.

## 3.2   Isostaticity and rigidity percolation

In the context of contractile active gels, we are interested how the connection probability $p$ affects the mechanical constitutive properties of the system, such as the shear modulus $G$. *Central force* models for marginal mechanical stability [58] offer a simple approach to understand linear elasticity. These resemble percolation models, except bonds represent mechanical springs rather than conduits. Shearing (or extending) the whole network causes deformation of individual springs, which respond with a linear stretch modulus $\mu$.

The onset of mechanical rigidity can be found by balancing mechanical degrees of freedom against constraints arising from network connectivity. There are $Nd$ degrees of freedom for the sites (or nodes) for a system with dimensionality $d$ and $N$ sites. The number of constraints due to pair-wise bonds is $\frac{1}{2}NZp$, with $Z$ the number of nearest-neighbor sites (e.g. $Z = 6$ for a 2D triangular lattice, or $Z = 2d$ for $d$-dimensional square lattices). These two quantities balance at a value of $p$ known as the *central-force isostatic point* $p_{CF} = \frac{2d}{Z}$. This point corresponds to the onset of mechanical rigidity or marginal stability of the network. For systems with only central-force interactions (e.g. random spring networks), $p_{CF}$ also corresponds to the *rigidity percolation threshold* $p_R$.

The rigidity threshold separates mechanical stability ($G > 0$, *elastic*) from instability ($G = 0$, *floppy*). Near $p_R$, the scaling relation $G \sim \mu a^2 (p - p_R)^{f_R}$ (with rigidity exponent $f_R \approx 1.4$ for 2D central-force models [167-170] and $a$ bond length) shows that $G$ increases as a power law in the distance of $p$ from $p_R$, similarly to the conductivity threshold. But the rigidity problem remains fundamentally different, because of the vector nature of the forces involved [171]. Indeed, the critical exponent $\nu_R \approx 1.16$ in 2D (Table 2) demonstrates that the central-force rigidity threshold represents a universality class distinct from the conductivity threshold [168,169,172]. The bonds inside the spanning cluster that bear stress (*rigidity backbone*) form a network that is more compact (higher fractal dimension) than the conduction backbone [173]. At $p_{CF}$, the rigidity backbone and shear modulus undergo a second-order transition, whereas the density of the stress-bearing cluster undergoes a first-order transition [174].

### 3.3  Fiber bending

Many real soft matter systems cannot be described as random spring networks [171]. Percolation models have been tailored to diverse material types including gels [175], foams [176], glasses [177], and granular systems [178]. The rigidity threshold can change when models include non-central-force interactions such as polymer branching [179], thermal/entropic effects [180,181], repulsive contacts [182], and interparticle friction [183]. In the context of cytoskeletal active gels, the most appropriate models are *fiber network models* [61,104,170,184-190]. These models treat filamentous networks as random networks of fibers with a finite bending rigidity $\kappa$ in addition to the stretch modulus $\mu$. Furthermore, hinge constraints may represent the presence of crosslinks connecting filaments. The finite bending modulus of fibers mechanically stabilizes fiber networks at connectivities below $p_{CF}$ and lowers the rigidity threshold $p_R$ [170,187,191,192] (Table 2). Thus sub-isostatic fiber networks ($p_R \leq p < p_{CF}$) have a finite shear modulus, which is proportional to $\kappa$ because the network modulus is dominated by non-affine bending modes. The rigidity exponent $f_R$ increases compared to its value for central-force models (Table 2). Networks with $p$ near $p_{CF}$ are predicted to exhibit diverging strain fluctuations, and $G$ depends on $\mu$ and $\kappa$ [61]. Above $p_{CF}$ networks stretch affinely, so the shear modulus is controlled by the filament stretch rigidity, $G \sim \mu$ (Fig. 3).

### 3.4  External driving

In addition to connectivity, *external driving* imposed by a mechanical deformation (using shear, extension, or compression) also affects the mechanical stability of a filamentous network. External stresses can either stabilize networks by providing a stabilizing field that decreases the rigidity threshold or destabilize networks by causing mechanical failure (Fig. **4**).

Bulk expansion and shear can stabilize networks that are initially below rigidity percolation and hence floppy. These *marginal networks* are only slightly underconstrained. For networks initially below the rigidity percolation transition, external strains pull out soft modes (deformation modes that cost no energy), so the network becomes elastic beyond a non-zero critical strain [193,194]. In fiber network models, external driving induces dramatic stiffening and a transition to a stretch-dominated elastic response at a critical shear strain [192]. This strain-controlled criticality can account quantitatively for the nonlinear mechanical response of athermal collagen networks and likely applies to other athermal fiber networks as well [191,192].

Conversely, external stresses can destabilize initially stable networks when they cause *failure*, or breaking of a material into disjoint pieces. Griffith's criterion provides a simple model of brittle failure of solids [195]. Let a homogeneous material with Young's modulus $E$ contain a single crack of length $a$. Failure depends on a balance between bond-breaking and free-surface energies. The material near the crack tips will fail when the applied stress exceeds a threshold value $\sigma_F \sim (\gamma E / a)^{1/2}$, where $\gamma$ denotes the surface tension. Once the material begins to fail, stresses redistribute and accumulate on the tips of the growing crack, producing more material failures that cause the crack to grow further. This positive feedback loop leads to an instability, where a straight crack propagates across the entire material.

Griffith's criterion neglects inhomogeneity, which prevails in disordered fiber networks [196]. Failure nucleation and propagation are sensitive to weaker sites in inhomogeneous systems; thus, experimentally measured values of $\sigma_F$ show sample-to-sample variability [197], and real failures leave a fractal crack [198]. Inhomogeneity is naturally accounted for in percolation models because of the stochastic nature of connections (reviewed in [199]). Furthermore, percolation models can include various kinds of bond-breaking rules [200-205], yielding predictions for $\sigma_F \sim (p - p_C)^{f_F}$ with $f_F$ the fracture exponent. One experimental study on metal plates with drilled holes found $f_F \approx 1.7$ in 2D; similar measurements for $E$ and $a$ yielded exponents that agreed with Griffith's criterion [206]. Percolation models can also account for the interaction of the stress fields from multiple flaws, since they already contain multiple sites where the network can fail [203,207]. These models have shown that the system fails by the coalescence of microcracks, which join and percolate across the system to form one large macroscopic crack. Failure percolation belongs to the same universality class as conductivity percolation [203].

Griffith's criterion also seems to hold for brittle-like failure in viscoelastic fluids [208]. However, it is unclear how well Griffith's criterion can describe failure of other transient networks including crosslinked actin networks, where rupture tends to be stochastic [209,210]. A more suitable alternative class of models to describe failure may be *fiber bundle models* (reviewed in [211]). These conceptually simpler models also predict the percolation of microcracks [212], as well as additional phenomena such as the failure-time distributions [213] that underlie the creep response in gels that precede failure [214]

Fig. **4** shows a conceptual diagram summarizing the mechanical response of passive filamentous networks as a function of network connectivity (bond probability $p$) and external driving. We identify three states: floppy networks (I) that are not connected enough to resist external loads, elastic networks (II) that are mechanically stable and store elastic energy, and ruptured networks (III) where a large crack breaks an elastic network

into two disconnected clusters. The boundaries are schematic, and the behavior near the junction of the three phases is poorly understood.

# 4   The phase behavior of active systems

So far we have seen how percolation models can be used to quantify the contribution of connectivity to the electrical and mechanical response of passive systems. In this section, we now focus on systems which are driven internally by molecular motor activity. Theoretical and experimental studies have identified a wide range of behaviors, including active flows, enhanced diffusion, directional transport, and contraction. We provide an overview of these studies, and attempt to unify a broad range of behavior in a proposed tentative state diagram (Fig. **5**). (Note that the terminology "state diagram" refers to regimes of non-equilibrium behavior or response to motor activity, rather than thermodynamically stable equilibrium phases or states.)

The axes of the state diagram are *network connectivity* and *motor activity*. In percolation models, network connectivity is given by the probability of either a bond forming between nearest-neighbor sites, or of a hinge constraint on a site with multiple bonds. Experimentally, network connectivity is some function of actin filament length, entanglement length, and crosslink concentration. Similarly, motor activity in experiment is some function of motor ATPase activity, duty ratio, and processivity. The functional dependencies mentioned above are not trivial, as detailed in §5.

The diagram we propose comprises four main regimes of network response to motor activity: *active solutions* (I), *prestressed gels* (II), *global contraction* (III), and *local contraction* (IV). The first three regimes are in analogy to passive systems. The fourth regime, in contrast, is special to active systems. The regimes are delimited by four boundaries: stress percolation (bold line), failure percolation (striped line), strain percolation (yellow line), and coarsening (dotted line). In the following, we describe the regimes and their boundaries.

## 4.1   Active solutions (I)

The primary interaction between actin filaments and myosin bipolar filaments is thought to be sliding. For systems that are weakly connected, the absence of constraints allows motor-filament sliding to proceed freely. In this *active solutions* regime (Fig. **5**, regime I), a variety of fluid-like phenomena have been reported. If myosins are tethered to a surface in so-called motility assays [215], dense suspensions of actin filaments glide with collective motions resembling bird flocks [14,216]. Similar behavior can be found in microtubule-kinesin motility assays [15]. If instead myosin and actin are free to move in solution, experimental measurements have shown that myosin activity decreases the apparent viscosity of the solution [217-219]. Theoretical models describe this behavior with an increased effective temperature [220-222], indicating that the energy released from myosin's ATPase activity is largely dissipated. In the presence of bundling proteins, myosin motors have been shown to induce super-diffusive collective transport of needle-like actin bundles [39,223]. Microtubule-kinesin solutions can display actively flowing nematic states [12,224], and swarms of bacteria give rise to apparent turbulence at low Reynolds number [13].

## 4.2  Stress percolation and prestressed gels (II)

In passive systems, mechanical stability is achieved at the rigidity percolation threshold, when mechanical constraints balance internal degrees of freedom. In active systems, mechanical constraints act to impede myosin-driven sliding. Once motors pull out slack in an underconstrained system, sliding ceases and motors exert ~pN stall forces [225], which subject the system to a tension-bearing state we call *prestressed gels* (Fig. **5**, regime II). At the boundary of this regime, mechanical constraints balance sliding compliance, allowing *stress percolation* (Fig. **5**, bold line) from motors across the gel. "Stress percolation" refers to the stabilization of floppy networks by motor stresses that act across the network [61,193,226]. This boundary resembles the rigidity percolation threshold in passive systems.

Direct experimental evidence for motor-driven, prestressed gels comes from optical tweezer microrheology studies of crosslinked actin-myosin gels revealing tension in the actin filaments [139,227], and from macroscopic measurements of the contractile stress exerted on the gel interface [38]. Similar tension-bearing states were observed in bundle contraction assays [228] and contractile ring assays [229] when anchored to soft gels. Prestressed gels appear homogeneous on macroscopic length scales. They may even appear homogeneous on microscopic scales, though experimental studies have also reported heterogeneous structure in certain cases, such as thick bundles in networks crosslinked with α-actinin, fascin, or filamin [38,227,230]. However, the density instabilities that characterize regime III (see below) are mostly absent in this regime.

It is interesting to note that motors' stabilizing effect in regime II contrasts motors' fluidizing effect in regime I. In order to further characterize the behavior at the stress percolation boundary, it will be interesting to develop more microscopic models that account for filament sliding, arrest by crosslinks, and motor-induced stabilization. Fiber network models coarse-grain out motor sliding, treating motor activity as effective contractile force dipoles [138,149], and thus likely break down at the stress percolation boundary [42,61,63]. Microscopic models that consider motor sliding [35,47,54,231] could be combined with constraint-counting arguments to investigate how motor-induced sliding interacts with soft modes.

Although prestressed gels do not contract, the contractile stresses they generate still serve important biological functions. In the actin cortex, motors maintain a constant level of prestress that sets the cell surface rigidity [232]. When cells prepare for cell division, additional myosin is recruited to the cortex, leading to an increased cortical tension and cell rounding [233]. During cytokinesis, motor activity in the cortex is differentially regulated between the cell poles and cell equator to ensure proper cytokinesis [234].

### 4.3 Failure percolation and global contraction (III)

If motor forces exceed the unbinding threshold of the crosslinks that maintain network stability, crosslink constraints can fail across the system (*failure percolation*; Fig. **5**, striped line) and motors are free to slide actin filaments. Well-connected actin-myosin gels respond to motor sliding by undergoing a *global contraction* (Fig. **5**, regime III), where the protein meshwork is collapsed to a tightly packed cluster with a higher density than the surrounding fluid. Global contraction events in reconstituted actin-myosin networks are visible with low-power microscopes and even the naked eye, which facilitated early experimental studies on the mechanisms of contractility in purified protein preparations [36,217,235]. In the absence of surface adhesion, globally contracting gels retain the shape of their container as they shrink uniformly [38,42,236]. This mechanically unstable behavior (*density instability*) is captured in active gel models [56,57,220,237] and contrasts the mechanically stable behavior of prestressed gels.

Experimental studies have identified the existence of a threshold myosin concentration above which contraction can occur. Two studies [36,38] found a threshold value of $R_M \approx$ 0.005 for actin gels, with $R_M$ the monomeric molar ratio of myosin to actin. Increasing the motor concentration past this threshold increased the contraction speed of the active gels [36,38]. What prevents gels with lower concentrations of myosin from contracting? Motor activity can enhance network elasticity [61,68,138,227,238], which may stabilize networks against motor-induced fracture. Furthermore, at high loads, muscle and non-muscle myosin II motors exhibit catch bond behavior [239-241]. Boundary adhesion may also need to be overcome. If the gel is strongly anchored to rigid boundaries, higher levels of contractile stress can accumulate across the gel. Failure could occur by either detachment from the boundaries [242] or the formation of large, microscopic cracks in the bulk [42].

Experimental studies have also found that global contractions occur above a threshold connectivity, either as a minimum crosslink concentration (Table 3) or minimum actin concentration ($c_A \approx 7.5 \text{ μM}$ [39], $c_A \approx 12 \text{ μM}$ [243], both measured for F-actin in the absence of proteins that regulate filament length). The minimum crosslink concentrations are close to the minimum crosslink concentrations required for gelation in the absence of motors (cf. §2.2). In the absence of crosslink proteins, contraction is usually not observed, but there are exceptions. When the ATP concentration becomes sufficiently low, motors themselves strongly bind actin and can cause contraction (or *superprecipitation*) [244]. Also, when the pH falls below 6, myosin binds more strongly to actin and promotes contractility [245]. In 1D tethered bundles [228] and 2D systems [41,148], contractions can occur in the absence of crosslink proteins, provided the myosin concentration exceeds a threshold value. Below this value, motors slide across more stationary actin filaments [246]. Above this value, myosin and actin move together in a velocity field with a negative divergence [246], indicating a density instability. How can 1D and 2D systems contract without added crosslinks? The connectivity threshold at the isostatic point depends linearly on the spatial dimensionality. Similar scaling likely holds for stress percolation and failure percolation in contractile active gels. Furthermore, the role of filament entanglements is amplified in lower dimensions [96].

In the limit of high connectivity, one would expect the network's larger elastic modulus to resist motor stresses and inhibit contraction [57]. Indeed, this expectation agrees with some experimental studies. Filamin-crosslinked gels contract more slowly when either more crosslinks are added or when actin filaments are longer [36]. α-actinin-crosslinked gels do not appear to contract above a threshold crosslink concentration ($R_X \approx$ 0.2) [38]. However, this behavior has not been observed for all crosslink types (see §5.2 below).

Global contraction events resemble syneresis or sintering events in polymer gels [247-250]. But crosslinked actin-myosin gels differ from most polymer gels due to the enzymatic activity of myosin molecular motors. As contraction proceeds, myosin bends, severs, and disassembles actin [41,66,67,236,251], rendering contraction events irreversible. Contraction events may not necessarily consume all protein, and may leave behind a sparser network of proteins, which may initiate a subsequent contraction wave [252]. Actin-myosin gels in cell extracts can even contract multiple times in waves or in a more continuous steady state [38,253,254], likely due to the presence of proteins that facilitate actin turnover.

Contractile events in cells and developing tissues require coordination of myosin activity over long length scales. If the contractile network is unanchored and free to contract, the resulting contractile strains could be used for intracellular transport, as has been suggested for chromosome congression in starfish oocytes [23]. Alternatively, if the network is well anchored to cell membranes and cells are free to deform, contractile activity can drive cell and tissue shape changes, as has been observed with cells and collagen tissues that invaginate from adhesion sites [255-258]. Furthermore, cells build transcellular actomyosin networks with a sarcomeric-like ordered arrangement of actin and myosin filaments to coordinate contraction across some epithelial tissues [259]. Contraction events are also regulated by biochemical signaling and, in turn, contribute to signaling through biomechanical feedback [22]. Regulation is likely needed to temper myosin activity; otherwise excessive stresses could rip the network apart into several disjoint clusters (see regime IV below). This phenomenon has been observed in developing *Drosophila* mutants with reduced cell-cell adhesions, where the ventral furrow rips apart into clusters of cells during gastrulation [260]. Smaller ruptures and subsequent repair by zyxin-mediated pathways occur in intact actomyosin structures such as stress fibers [261,262]. The tissue-scale contractions found in gastrulating embryos do not proceed as continuous, spatially uniform contractile events; rather, several pulsatile bursts of contractile activity that span single-cell length scales drive contraction [263-265]. These pulses are regulated by pulses of myosin phosphorylation [266] which arise from an interplay between feedback and self-organization [267-270].

## 4.4   Coarsening and local contraction (IV)

Contraction events do not necessarily span system length scales. Many experimental studies have also reported *local contractions* (Fig. **5**, regime IV). Here, motors compact crosslinked actin filaments into many small clusters, which tend to comprise a myosin core and an actin coat [37,40,41,236]. These clusters are usually disordered, though more ordered rings and asters can occur [230]. Local contractions may also occur when excessive motor activity breaks an initially prestressed gel into many clusters [42,260].

There are several differences between global and local contractions. Local contraction events tend to occur above threshold crosslink concentrations that are up to an order of magnitude lower than those for gelation in passive systems or global contractions in active systems (Table 3). Local contraction events often occur over a certain period of time after network formation and then cease [41,42,66,236]. However, there may still be a mechanically stable background network between compacted clusters. This intervening network may allow motors to actively coalesce nearby clusters over longer time scales [40,271] or even disassemble clusters in an apparently reversible fashion [37]. This type of myosin contractile activity can manifest itself in marked nonthermal fluctuations of inert probe particles embedded in the network [139], which may contribute to nonthermal fluctuations observed in cells [1,272-275]. Myosin has also been observed to nucleate actin-filament polymerization in the presence of fascin [236]. Many of these hallmarks resemble the *coarsening* (Fig. **5**, dotted line) or aggregation behavior of phase-separating soft-matter systems [276-278].

Continuum models of active systems have characterized the onset of ordered states that resemble local contractions. These states are characterized by a density instability with broken spatial symmetry and an apparently well-defined length scale [53,279,280]. Experimental studies have identified different ways to determine the length scale of contractions, by either measuring the size of contracted clusters [236,281], measuring the average distance between contracted clusters [41], or tracking cluster expansion in time-reversed movies [42]. The contraction length scale was shown to increase with increasing connectivity, achieved either by varying the crosslink concentration [42] the actin filament concentration [39], or actin filament length [281]. The effect of motor activity, however, is not as clear (see §5.1 below). Another physical effect on contraction length scale is surface adhesion, which can attenuate strain propagation [41]. Friction of the actin-myosin cortex with the membrane and/or cytosol reduces the hydrodynamic length scale of myosin-driven cortical flows in *C. elegans* embryos [282] or could result in negative stiffness [283]. Additionally, the geometry of actin-filament nucleation can strongly affect the spatial organization of actin filaments, and thus contractility [43,284].

There are many *in-vivo* observations of small and dense myosin clusters, or *foci*, resembling those observed in reconstituted actin-myosin networks, including the cortex of developing *C. elegans* embryos and the cell equator of mitotic cells [282,285]. However, the length scales observed in cells are likely controlled by reciprocal feedback between contraction and regulatory reaction-diffusion systems [286]. Interestingly, contraction of the actin-myosin cortex is thought to affect the spatial distribution and possibly downstream signaling of lipid-tethered proteins [281,287].

## 4.5    Strain percolation

Both the global and local contraction regimes are characterized by active, contractile strain generation that occurs on long and short length scales, respectively. What is the behavior between these two regimes, at the boundary we refer to as contractile *strain percolation* (Fig. **5**, bold yellow line)? In studies of quasi-2D fascin-crosslinked gels we found evidence for a "critically connected" state near this boundary, where the network breaks up into clusters with a power-law size distribution [42]. The experimentally measured Fisher exponent $\tau \approx 1.91 \pm 0.06$ of the cluster size distribution is close to the conductivity percolation threshold, which predicts $\tau = \frac{187}{91} \approx 2.05$ in 2D (cf. §3.1). This agreement would suggest that the conductivity transition underlies contractile strain percolation. The conductivity transition governs material failure by the formation of microcracks which join to form one macroscopic crack, which breaks passive systems into two large clusters according to Griffith's criterion (cf. §3.4). In contrast for contractile active gels, multiple cracks are nucleated concurrently, which allows for clusters with a power-law size distribution [42]. This behavior is likely only possible for internally driven active systems, which generate stresses within the entire material. Furthermore, crosslinked actin gels tend to fail by crosslink unbinding (failure of nodes) rather than actin filament breakage (failure of bonds). The combination of internal loading and failure by crosslink unbinding results in a material where stresses do not necessarily accumulate near failures [288].

Furthermore, recent models have found further evidence that contractile strain percolation is distinct from conductivity percolation. These models arose from the observation that the hyperscaling relation and the fractal shape of clusters at the conductivity percolation threshold implies $\tau \geq 2$. This condition appears inconsistent with the experimentally measured value of $\tau \approx 1.91$ for contractile networks near the strain percolation threshold. To resolve this inconsistency, attention has turned to the role of enclaves. Presumably, larger clusters engulf enclaves during contraction due to steric interactions, resulting in Euclidean, rather than fractal clusters [288,289]. In particular, Ref. [289] predicts that a "no-enclaves percolation" (NEP) model characterizes a transition with a novel universality class, which allows $\tau < 2$ and predicts $\tau \approx 1.82$. Furthermore, the transition is of mixed-order, being second-order in the correlation length but first-order in the order parameter. These claims are currently under debate [242,290-292]. It is interesting to note that the rigidity percolation transition for passive systems is also of mixed order [174]. Additionally, the NEP model can be mapped to the problem of holes in the percolation backbone, which yields $\tau = 187/96 \approx 1.948$ [292]. Future studies which investigate other critical exponents, such as the correlation length exponent $v$, could determine whether the strain percolation transition corresponds to a novel universality class.

Critical behavior is usually only observed in a very narrow region of phase space. Experimentalists must finely tune their system and bring it to this narrow region in order to observe critical behavior. However, the critically connected state we reported for actin-fascin-myosin gels [42] occurred over a surprisingly broad region of crosslink concentration. Hence, strain percolation is "robust" in the sense that networks whose initial connectivities exceed the critical transition may still be brought down to a critical state. This property bears some similarity to models of "self-organized criticality" [293], though these models rather tend to describe dynamic fluctuations which drive a system to a specific critical point.

Intriguingly, power-law distributions have been observed for inter-cluster distances in 1D contractile bundle assays [294]. The observed distribution exponent of 1.51 may possibly relate to critical behavior near the 1D strain percolation transition.

# 5  Microscopic contraction mechanisms

We have identified different classes of behavior in active systems: fluid-like sliding motions (I); mechanically stable, tension-bearing gels (II); and motor-driven contractions of varying length scales (III and IV). These behaviors are all attainable in experiments using actin, myosin, and crosslink proteins as minimal components. The diagram we propose organizes these behaviors according to two abstract quantities: motor activity and connectivity. This approach attempts to assemble a broad, unified picture. However, a more microscopic picture requires concrete understanding of the proteins' molecular properties. Below, we summarize current knowledge of these properties and their relation to connectivity and motor activity.

## 5.1  Motor activity and myosin biochemistry

Myosin ATPase activity depends on several biochemical parameters (reviewed in [295]), including *duty ratio* (fraction of the ATPase cycle where the motor tightly binds F-actin), *processivity* (number of successive cycles before the motor diffuses away from the filament), and *velocity*. Duty ratio and processivity are small for individual skeletal muscle myosin II motors [296]. But once assembled into bipolar filaments, the effective duty ratio and processivity increase, allowing the ensemble of motors to slide actin filaments [297-299]. This behavior contrasts with the directed, processive transport mediated by other myosins such as myosin VI and by most microtubule-associated kinesin and dynein motors. We note that dimers of myosin VI exhibit contractile behavior in micropatterned ring assays [229,284], though in cells they are mainly thought to mediate intracellular transport [300].

The effective motor duty ratio and velocity of myosin II bipolar filaments depend on the individual myosin's properties and motor filament. In humans, the myosin II class includes skeletal muscle myosin II, smooth muscle myosin II, and three nonmuscle myosin II isoforms known as nonmuscle myosin IIA, IIB, and IIC [301]. The myosin II motors which are expressed in skeletal and smooth muscle are densely packed into long bipolar filaments (*thick filaments*) that are surrounded by a dense linear array of actin filaments. The thick filaments contain ~300 myosins and are ~2 μm in length [302]. Smooth muscle myosin II forms side polar filaments of ~176 molecules that are ~0.6 μm in length [303], though filament length can change during activation [304,305]. This property is thought to underlie the structural malleability of smooth-muscle cells [306]. Myosin II motors of non-muscle cells form much shorter bipolar filaments (often referred to as *minifilaments*) of ~30 myosins that are close to 300 nm in length [137,259,307-311]. These bipolar filaments are either embedded in actin-based contractile bundles such as stress fibers or supracellular junctional belts [312], or in disordered meshworks such as the actin cortex [313]. Tethered bundle assays have suggested that myosin filament length governs contractility to a greater extent than the specific myosin II isoform [314]. Regulation of the phosphorylation state of non-muscle myosin II by a set of kinases and phosphatases controls both the ATPase activity and bipolar filament assembly, providing cells with the means to spatiotemporally control contractility. Regulatory kinases and phosphatases, in turn, are themselves regulated by the Rho signaling pathway, which acts as an intermediary between external signals and direct actomyosin control [315,316].

In reconstitution assays, the activity of myosin II is usually tuned by varying either the biochemical buffer conditions or the total myosin concentration. Experiments are often performed at low-salt and low-ATP (50 mM KCl, 0.1 mM ATP, and pH 7.4) to promote a high duty ratio. Higher KCl concentrations decrease the binding affinity of skeletal myosin II to actin filaments [317] to the point where they prevent sliding of actin filaments in gliding assays above 60–100 mM of KCl [318]. Furthermore, increasing the monovalent salt concentration typically decreases the size of the bipolar filaments formed by purified skeletal muscle myosin II to 10-20 at around 100–150 mM monovalent salt [314,319-325]. Higher concentrations of ATP do not greatly affect the filament sliding speed [326,327], but motors spend more time in a weakly-bound state, thus reducing processivity [299,328]. The contractile activity of 2D gels appears to be maximal in a window around 0.1 to 1 mM [66]. Solutions of driven needle-like actin bundle clusters find a maximum in collective pulsatile behavior for ATP concentrations near 0.1 mM [223]. When the pH is reduced from physiological values slightly above pH 7 to pH 6.4 or less, ADP release and thus the ATPase cycle slow down and motors remain strongly bound to actin filaments for longer periods of time, likely increasing connectivity [245,329].

Biochemical conditions thus determine contractility. Changing the motor activity can affect the contraction length scale. This was observed in studies of critically connected fascin-crosslinked gels, where decreased motor concentration [42], or increased KCl or ATP concentrations [243] increased the contraction length scale. Meanwhile, studies of locally contracting gels found smaller clusters with decreased myosin activity achieved by either increasing the KCl concentration [236] or decreasing the myosin concentration [40]. Similarly, tethered bundle assays found that added KCl decreases contractile strain rate and tension [314].

## 5.2   Network connectivity and crosslinks

Network connectivity can be modulated experimentally by varying the concentration or length of actin filaments, as well as the crosslink concentration and type. Connectivity appears to affect myosin-driven contractility non-monotonically. This non-monotonicity was recently investigated in a study of micropatterned contractile rings [43]: at low connectivity, adding connections allows motors to contract coherently over longer length scales; at the strain percolation threshold, there is a maximum in contractility (tension or strain rate); excessive crosslinking past this maximum inhibits buckling and thus contractility. Indeed, high amounts of α-actinin or filamin crosslinks appear to inhibit contraction [36,38]. However, studies on gels crosslinked with fascin or cortexillin showed no evidence of slower or inhibited contraction at high crosslink concentration [42,252]. In fact, mixing fascin and cortexillin in the same gel can yield faster contractions than gels with fascin or cortexillin alone [252]. These results suggest that the response of crosslinked gels to motor activity sensitively depends on how crosslinks bind and connect actin filaments.

Crosslink-mediated bundling can affect contractility in active gels in different ways. On the one hand, bundles should exert stronger forces than isotropic meshworks [330,331]. On the other hand, stiff bundles reduce buckling and can inhibit contraction. For example, one study of actin-fascin systems found that motors can cause superdiffusive transport of clusters of driven, needle-like bundles, rather than contraction [39]. Similarly, oligomers of kinesin motors slide stiff microtubule bundles without contraction [12]. However, contraction may still be possible in the absence of buckling, as shown for microtubule-dynein systems [44,153].

In addition to bundling, the effect of bundle polarity adds further nuances to the contribution of crosslinks to contractility. Studies of micropatterned contractile rings have demonstrated that myosin contractility is maximized when actin filaments are antiparallel [43,284]. In actin-bundle arrays, motors move towards regions of low polarity [228,294], reminiscent of the polarity sorting found in fibroblasts [332]. Many studies of contractile active gels have used the crosslink protein fascin. In passive systems, fascin forms unipolar bundles [333] that form by a fast zippering process [334]. Fascin binds actin filaments with a narrow angle [98], perhaps as a consequence of its small size [107] and potentially its unique β-trefoil domains [95]. In active gels, fascin's polar binding property is thought to be essential for the formation of self-organized ordered structures [230] or nucleation of actin filaments on myosin [236]. However, fluorescence microscopy shows that structure on the 1–10 μm scale can vary from bundled to unbundled, depending on the type of crosslink [252] or on monovalent salt concentration [236,243]. It is possible that assembly kinetics, strong motor stresses, or actin-filament entanglements inhibit the formation of parallel bundles in contractile active gels.

## 5.3   Interplay between motor activity and connectivity

One further microscopic mechanism needs careful consideration: motor activity and connectivity can affect each other. On the one hand, motor activity can reduce connectivity by a number of mechanisms. First, stresses due to motor activity increase tension within crosslinks and will generally tend to increase the crosslink dissociation rate (slip bond behavior) [64]. Forced crosslink unbinding results in an attenuated length scale of contractile strain propagation [42]. Myosin activity can also sever actin filaments during contraction [41,66], depolymerize them [67], and disassemble larger structures [67,230,236]. On the other hand, motor activity can increase connectivity. Myosin motors themselves increase connectivity by acting as (transient) crosslinks between actin filaments. Moreover, many myosins (skeletal muscle, smooth muscle, and non-muscle myosin IIB) exhibit catch bond behavior. Single-molecule measurements with optical tweezers showed that the binding affinity of myosin motors to actin increases with applied force, for forces up to ~6 pN [239,240,335]. This catch-bond behavior may promote contractility [336]. Furthermore, motor activity can mechanically stabilize low-connectivity networks by pulling out slack [61,62]. Although this effect does not directly contribute to connectivity, it may affect stress redistribution through crosslinks in the network. In addition to these mechanisms, myosin motors may also directly interact with certain crosslinks, perhaps to aid in mechanosensing. During cytokinesis, myosin has been shown to bind cortexillin in *Dictyostelium* cells [337] and anillin in *Drosophila melanogaster* and human cells [338]. Furthermore, cortical tension can be regulated by cortical thickness, as well as actin-filament length regulators [21]. The relationship between motor activity and connectivity in experimental systems is therefore rich, and should be considered when modeling contractile active gels.

# 6  Summary and outlook

Contractile active gels are unique materials which spatially integrate microscopic stresses to power macroscopic behavior. Experimental and theoretical studies of actin-myosin systems have revealed that contractility is determined by an interplay of two physical parameters: network connectivity and motor activity. We proposed a state diagram that unites a large body of published experimental data and theoretical/computational predictions in terms of these two parameters. We expect that the behavior described here is generic and applies to contractile active gels comprising non-cytoskeletal components. For instance, a recent study successfully developed a non-cytoskeletal active gel composed of linked DNA tubes and processive motors [339]. This gel exhibited nonthermal fluctuations similar to those found in actin-myosin systems [139,340,341]. Advances in DNA nanotechnology and synthetic chemistry will likely provide more examples of synthetic active gels in the future [18,342,343].

Purified acto-myosin networks provide a powerful model system to study the physical principles that govern contractility in cells. However, we should emphasize that the state diagram proposed in this review is specific to actin-myosin networks in which actin turnover is slow. In cells, there is continuous rapid turnover of actin, which acts in conjunction with motor activity [48,76,344-346]. Physical parameters regulate actomyosin contractility in conjunction with biochemical regulation, so that the same actin and myosin components can be used to build different types of contractile structures to power diverse physiological processes [347]. For instance, recent observations of *Dictyostelium* amoebae showed that a cooperative interaction between motors and crosslinks aids mechanosensation [337,348,349]. During cytokinesis, cortical flows deliver myosin II to the cell equator, which contributes to local contractile ring formation in concert with biochemically regulated myosin activation and filament formation [350]. Large-scale cortical flows have also been shown to aid segregation of membrane-bound cell polarity factors in embryos [351], and local contractions of the actin-myosin cortex in hamster ovary cells have been shown to cause clustering of cell surface proteins that are involved in cell signalling [287,352]. These observations suggest a strong link between the physical properties of active gels and the regulation of the plasma membrane [281]. Such a link could allow for a direct role for contractile active gels in the regulation of biochemical signaling networks [286,351,353].

# 7 Acknowledgements

This work is part of the research programme of the Netherlands Organisation for Scientific Research (NWO). G.K. acknowledges support from a Starting Grant from the European Research Council under the European Union's Seventh Framework Programme (FP/2007-2013) / ERC Grant Agreement n. [335672]. J.A. acknowledges support from the U. S. Army Research Laboratory and the U. S. Army Research Office under grant number W911NF-14-1-0396, as well as the National Science Foundation under Grant No. NSF PHY11-25915. F.C.M. was supported in part by the National Science Foundation (Grant PHY-1427654). We furthermore thank Shiladitya Banerjee, Chiu Fan Lee, Gunnar Pruessner, Margaret Gardel, Martin Lenz, Michael Murrell, Thibaut Divoux, and Stephan Grill for insightful comments and discussions.

# 8 Figures and tables

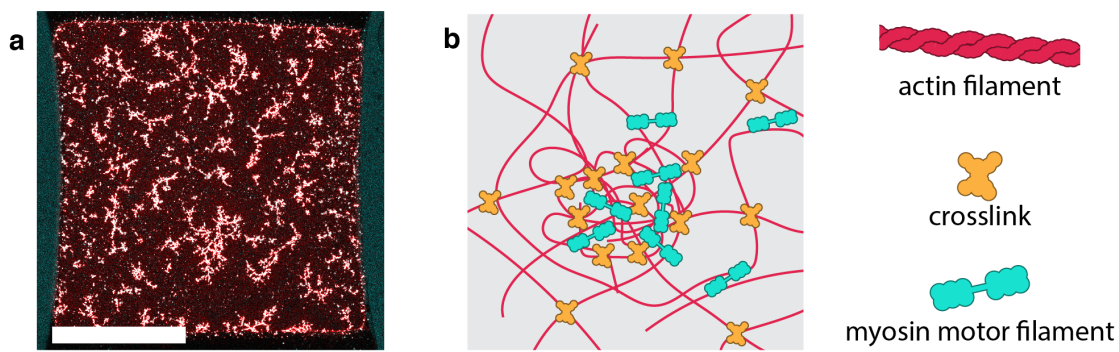

**Figure 1.** Contractile active gels comprise three main ingredients. **a**. Fluorescence micrograph of a contractile active gel of actin filaments, crosslinks, and myosin motor filaments. Scale bar 1 mm. **b**. Schematic depicting the three main ingredients: actin filaments (red lines), crosslinks (orange crosses), and myosin motors (cyan dumbbells).

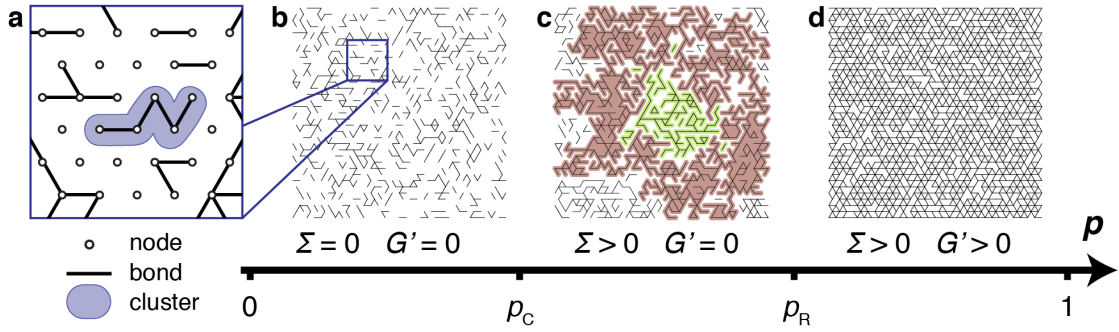

**Figure 2.** Bond probability $p$ in random networks determines network connectivity, cluster shape, and constitutive properties, e.g. conductivity $\Sigma$ and storage modulus $G'$. **a**. Close-up schematic of a bond percolation model with a 2D triangular lattice of nodes (circles), bonds (lines), and clusters of connected bonds (one example of a cluster highlighted in blue). **b**. Network ($p = 0.2$) comprising small, disjointed clusters. **c**. Network ($p = 0.35$) with a fractal spanning cluster (red) that surrounds enclaves (light green). **d**. Network ($p = 0.8$) comprising one solid spanning cluster. Networks above the conductivity threshold $p_C$ have finite $\Sigma$. Networks above the rigidity threshold $p_R$ have finite $G'$.

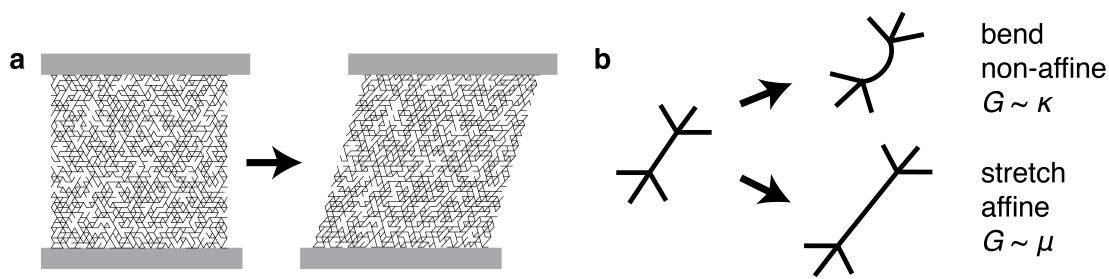

Figure 3. a. Schematic of a fiber network deforming affinely in response to an externally imposed shear. b. Illustration of a network segment undergoing two kinds of deformation. When bending deformations dominate (top), segments do not deform affinely. The network's mechanical modulus depends on the filament bend modulus $\kappa$. Meanwhile, stretching deformations (bottom) produce affine deformations. The network's mechanical modulus depends on the filament stretch modulus $\mu$. For semiflexible filaments, like actin, the stretch modulus results from pulling out entropic fluctuations.

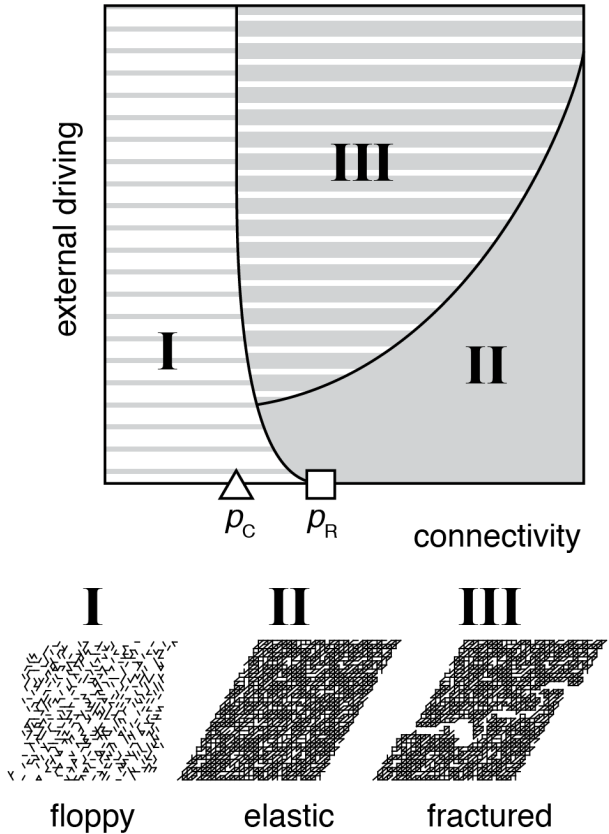

**Figure 4.** Conceptual state diagram of passive systems, with three regimes of mechanical response as a function of connectivity and external driving (stresses or strains): floppy materials, which have a zero storage modulus (I); elastic materials, which are mechanically stable (II); and fractured materials, which fail by the formation of a large crack (III). Depicted along the connectivity axis are the conductivity threshold $p_C$ (triangle) and rigidity threshold $p_R$ (square).

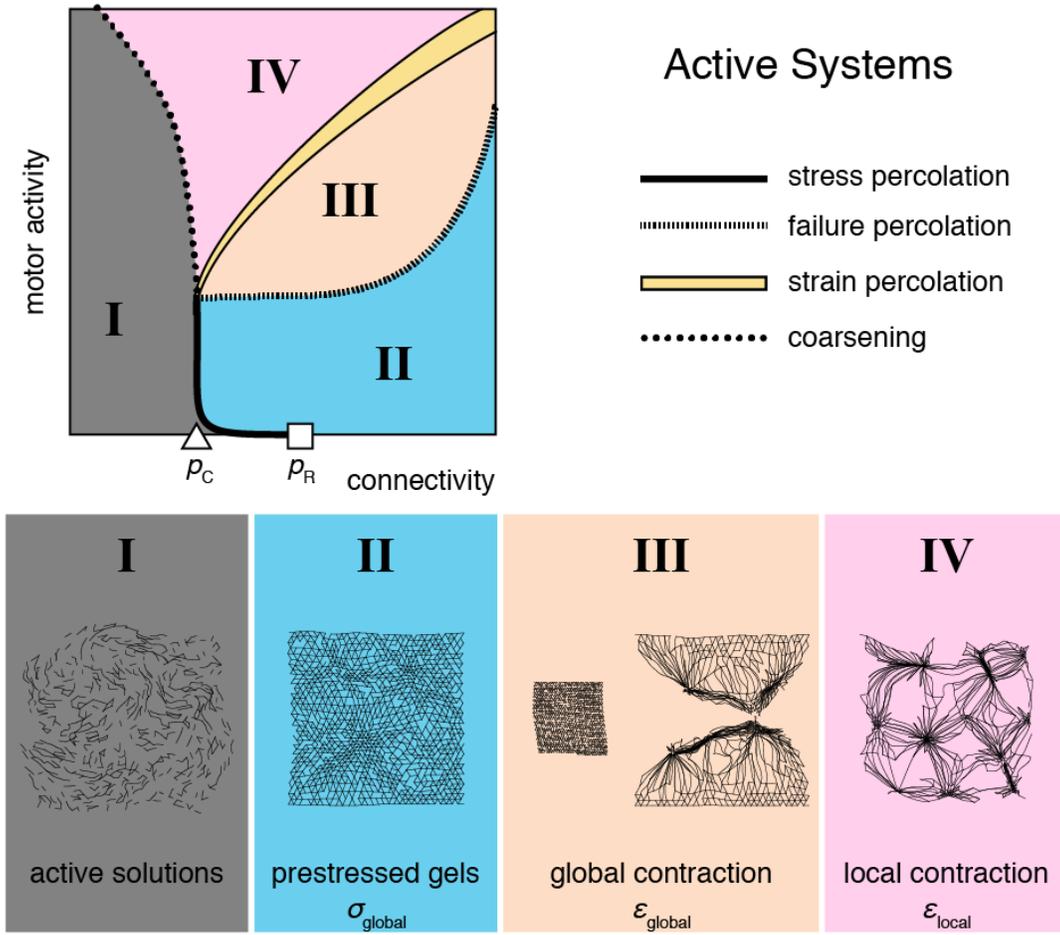

**Figure 5.** Proposed conceptual state diagram of active systems, depicting four regimes of mechanical response to internal driving from motor activity. Active solutions (I) exhibit fluid-like motions. Prestressed gels (II) are mechanically stable under motor loads and maintain contractile stresses across the gel. In global contraction (III), motors strain the network across system length scales by uniformly compacting unanchored gels into a large cluster, or by breaking anchored gels into large clusters. In local contraction (IV), motors compact gels into smaller clusters. There are four boundaries between the regimes. At stress percolation (solid black line), motor stresses balance mechanical constraints and percolate across the network. At failure percolation (striped black line), motor stresses overwhelm the network's mechanical constraints and cause contraction. At strain percolation (thick yellow line), the length scale of coherent contractile strain diverges. At coarsening (dotted line), motors begin to compact the network around them.

| Crosslink | Onset of stiffness |
|-----------|--------------------|
| fascin | 0.01 [i) 354] |
| filamin | 0.001-0.01 [ii) 355] |
| α-actinin | 0.01 [i) 106] |
| scruin | 0.03 [iii) 110] |

**Table 1**. Values of the crosslink-to actin molarity ratio $R_X$ where the onset of stiffness occurs. Molarities of actin: i) 9.5 μM, ii) 24 μM iii) 12 μM.

| Quantity | Definition | Numerical value in 2D |
|----------|-----------|----------------------|
| Conductivity percolation threshold $p_C$ | Infimum value of $p$ which yields a system-spanning cluster [159] | $\approx 0.347$ [i) 159] |
| Fractal dimension $d_f$ | $s \sim l^{d_f}$ [iii) 159] | $= 91/48 \approx 1.90$ [ii) 159] |
| Fisher exponent $\tau$ | $P(s) \sim s^{-\tau}$ [iii) 159,165] | $= 187/91 \approx 2.05$ [ii) 159] |
| Conductivity exponent $f_C$ | $\Sigma \sim \Sigma_0(p - p_C)^{f_C}$ [iv) 159] | $\approx 1.3$ [ii) 159,166] |
| Correlation length exponent $\nu_C$ | $\xi \sim |p - p_C|^{-\nu_C}$ [iii) 159] | $= 4/3 \approx 1.33$ [ii) 159] |
| Susceptibility exponent $\gamma$ | $S' \sim |p - p_C|^{-\gamma}$ [iii) 159] | $\approx 43/18 \approx 2.39$ [ii) 159] |
| Central force isostatic threshold $p_{CF}$ | The value of $p$ where constraints balance degrees of freedom [58] | $= 2/3 \approx 0.667$ [v) 58] |
| Rigidity threshold $p_R$ | Infimum value of $p$ which yields finite elasticity [58,170] | $= 2/3 \approx 0.667$ [v) 58] $\approx 0.445$ [vi) 170] |
| Rigidity exponent $f_R$ | $G \sim \mu a^2(p - p_R)^{f_R}$ [vii) 170] | $\approx 1.4$ [viii) 170] $\approx 3.2$ [ix) 170] |
| Correlation length exponent $\nu_R$ | $\xi \sim |p - p_R|^{-\nu_R}$ [170] | $\approx 1.16$ [viii) 168,169,172] $\approx 1.3$ [ix)] |
| Fracture exponent $f_F$ | $\sigma_F \sim (p - p_F)^{f_F}$ [x) 206] | $\approx 1.7$ [x) 206] |

**Table 2.** Quantities which relate to conductivity, rigidity, and fracture percolation. i) Valid for the 2D triangular bond lattice. ii) Valid for 2D models irrespective of lattice. iii) Valid for $p \approx p_C$. iv) Valid for $p \gtrsim p_C$. v) Valid for central force models with the 2D triangular bond lattice. vi) Valid for fiber bending models with the 2D triangular bond lattice. vii) Valid for $p \gtrsim p_R$. viii) Valid for 2D central force models respective of lattice. ix) Valid for 2D fiber bending models irrespective of lattice. x) Valid for models with bond-breaking rules.

| Crosslink | Onset of global contraction | Onset of local contraction |
|---|---|---|
| fascin | 0.05 [i) iv) 42] | 0.002 [i) iv) 356] |
| filamin | 0.005 [ii) v) 36] | 0.0025 [ii) v) 36] |
| α-actinin | 0.05 [iii) vi) 38] | unknown |
| biotin-streptavidin | unknown | 0.001 [iii) v) 40] |

**Table 3**. Values of the crosslink-to-actin molarity ratio $R_X$ where the onsets of local contraction and global contraction occur. Molarities of actin: i) 12 µM, ii) 36 µM, iii) 24 µM. Myosin-to-actin molarity ratio $R_M$: iv) 0.01 v) 0.02 vi) 0.005.